\documentclass{ws-ijmpd}
\begin{document}
\markboth{G Govender, M Govender and K S Govinder} {Thermal behaviour of Euclidean stars}

\catchline{}{}{}{}{}
%
%%%%%%%%%%%%%%%%%%%%%%%%%%%%%%%%%%%%%%%%%%%%%%%%%%%%%%%%%%%%%%%%%%%%

\title{Thermal behaviour of Euclidean stars}

\author{G Govender\footnote{E-mail: 200301326@ukzn.ac.za}}
\author{M Govender \footnote{E-mail: govenderm43@ukzn.ac.za}}
\author{K S Govinder\footnote{E-mail: govinder@ukzn.ac.za}}

\address{Astrophysics and Cosmology Research Unit, School of Mathematical Sciences, \\
University of KwaZulu--Natal, Private Bag X54001, Durban 4000, South Africa}

\maketitle

\begin{history}
\received{Day Month Year}
\revised{Day Month Year}
%\accepted{Day Month Year}
%\comby{(xxxxxxxxxx)}
\end{history}

\begin{abstract}
A recent study of dissipative collapse considered a contracting sphere in which the areal and proper radii are equal throughout its evolution. The interior spacetime was matched to the exterior Vaidya spacetime which generated a temporal evolution equation at the boundary of the collapsing sphere. We present a solution of the boundary condition which allows the study of the gravitational and thermodynamical behaviour of this particular radiating model.

\end{abstract}


\section{Introduction}    %) A SECTION HEADING

The study of dissipative gravitational collapse achieved prominence with the presentation
of the junction conditions by Santos\cite{santos} in 1985. Earlier work on collapsing stars in general relativity
assumed the exterior spacetime to be empty and as a consequence, it was required  that the pressure at the boundary vanish.
Santos provided the general junction conditions required for the smooth matching of a spherically symmetric, shear-free spacetime to the exterior Vaidya\cite{vaidya} solution across a time-like hypersurface. An important consequence of the matching conditions is that the pressure on the boundary of the radiating star cannot be zero. It is assumed that the interior of the star is radiating energy in the form of a radial heat flux.
The junction conditions due to Santos rejuvenated the study of gravitational collapse and the end-states of radiating stars. The simplistic model of Oppenheimer and Snyder\cite{snyder} has been generalised to include pressure\cite{bon}, anisotropic stresses, electromagnetic field\cite{charge} and the cosmological constant\cite{thiru1}. These exact models, although simplified, give much insight into the dissipative collapse process as well as physical characteristics of the radiating star such as its temperature and luminosity.

What makes the study of dissipative collapse of stars particularly difficult is the solution to the boundary condition representing the conservation of momentum across the time-like hypersurface. While many exact solutions for shear-free radiating spheres have appeared in the recent literature, there are very few models that include the effects of shear in the interior of the star. One of the first exact models of a shearing radiating star that allowed for an analysis of the gravitational and thermodynamical behaviour of the stellar fluid was found by Naidu {\em et al}\cite{nol}. However, the model was restrictive in the sense that it was acceleration-free, but more importantly, the matter variables such as pressure and density became infinite at the center of the star. It was pointed out that this model could form part of a core-envelope model of a radiating star. Further exact shearing solutions were obtained by Rajah and Maharaj\cite{rajah} in which it was assumed that the particle trajectories within the stellar core were geodesics. An analysis of the temperature profiles for these models reveals unphysical behaviour  in that the temperatures closer to the surface of the star become negative.  A recent study of shearing, dissipative collapse considered a model of a spherically symmetric matter distribution in which the areal radius is equal to the proper radius throughout the stellar evolution\cite{her1}. These so-called Euclidean stars were shown to exhibit very interesting general properties. In this paper we present an exact solution to the boundary condition that determines the temporal evolution of a Euclidean star. Our solution allows us to study the physical and thermodynamical properties of this class of stars even when the stellar fluid is far from equilibrium. Since Euclidean stars are {\em not} acceleration-free we are able to draw comparisons with the earlier models of Naidu {\em et al}\cite{nol} and Rajah and Maharaj\cite{rajah}.

The basic structure of our paper is as follows: Following Herrera {\em et al} \cite{her1} we present the Einstein field equations for
the interior of the stellar fluid in section two. In section three we describe the exterior spacetime given by the Vaidya solution and we give the main junction conditions required for the smooth matching of the interior and exterior spacetimes. In section four we specify the Euclidean condition and a particular solution to the boundary condition. In section five we investigate the thermodynamical properties of our model and make comparisons to earlier work on shearing, radiating models of gravitational collapse.

\section{Shearing spacetimes}

The interior spacetime is described by the general spherically symmetric, shearing metric in comoving coordinates
\begin{equation}
ds^2 = -A^2dt^2 + B^2dr^2 + R^2(d\theta^2 +
\sin^2{\theta}d\phi^2) \, ,\label{metric}
\end{equation}
where $A = A(t,r)$, $B = B(t,r)$ and $R = R(t,r)$ are metric functions yet to be determined. The matter content for the interior is described by
\begin{equation} T_{\alpha\beta} = (\mu + P_{\perp})V_\alpha V_\beta + P_{\perp}g_{\alpha\beta} + (P_r - P_{\perp})\chi_\alpha\chi_\beta + q_\alpha V_\beta
          + q_\beta V_\alpha \,,\label{2}
\end{equation} where $\mu$ represents the energy density, $P_r$ the radial pressure, $P_{\perp}$ the tangential pressure and $q^{\alpha}$ the heat flux vector. The fluid four--velocity ${\bf V}$ is comoving and is given by
\begin{equation} V^\alpha = \displaystyle\frac{1}{A} \delta^{\alpha}_0 \,.
\label{2'}
\end{equation} The heat flow vector assumes the form
\begin{equation} q^\alpha = (0, q, 0, 0)\,, \label{2''}
\end{equation} since $ q^\alpha V_\alpha = 0 $ ensuring radial heat dissipation. We further have
\begin{equation}
\chi^{\alpha}\chi_{\alpha} = 1, \hspace{2cm} \chi^{\alpha}V_{\alpha} =0.\end{equation}
The expansion scalar and the fluid four acceleration are given by
\begin{equation}
\Theta = V^{\alpha}_{;\alpha}, \hspace{2cm} a_{\alpha} = V_{\alpha;\beta}V^{\beta},
\end{equation}
and the shear tensor by
\begin{equation}
\sigma_{\alpha_\beta} = V_{(\alpha;\beta)} + a_{(\alpha}V_{\beta )} - \frac{1}{3}\Theta (g_{\alpha \beta} + V_\alpha V_\beta). \end{equation}
For the comoving line element (\ref{metric}) the kinematical quantities take the following forms
\begin{eqnarray}
a_1 &=& \frac{A'}{A} \\
\Theta &=& \frac{1}{A}\left(\frac{\dot B}{B} + 2\frac{\dot R}{R}\right)\\
\sigma &=& \frac{1}{A}\left(\frac{\dot B}{B} - \frac{\dot
R}{R}\right),\end{eqnarray}
 where dots and primes denote differentiation with respect to $t$ and $r$ respectively. The nonzero
components of the Einstein's field equations for the line element
(\ref{metric}) and the energy-momentum (\ref{2}) are
\begin{eqnarray}
\mu &=& \frac{1}{A^2}\left(2\frac{\dot B}{B} + \frac{\dot R}{R}\right)\frac{\dot R}{R} - \frac{1}{B^2}\left[2\frac{R''}{R} + \left(\frac{R'}{R}\right)^2 - 2\frac{B'}{B}\frac{R'}{R} - \left(\frac{B}{R}\right)^2\right],  \label{14a} \\ \nonumber \\
P_r &=& -\frac{1}{A^2} \left[2\frac{{\ddot R}}{R} - \left(2\frac{\dot A}{A} - \frac{\dot R}{R}\right)\frac{\dot R}{R}\right]  + \frac{1}{B^2}\left(2\frac{A'}{A} + \frac{R'}{R}\right)\frac{R'}{R} - \frac{1}{R^2}, \label{14b}  \\
P_{\perp} &=& -\frac{1}{A^2}\left[\frac{\ddot B}{B} + \frac{{\ddot R}}{R} - \frac{\dot A}{A}\left(\frac{\dot B}{B} + \frac{\dot R}{R}\right) + \frac{\dot B}{B}\frac{\dot R}{R}\right]  \nonumber \\&&+ \frac{1}{B^2}\left[\frac{A''}{A} + \frac{R''}{R} - \frac{A'}{A}\frac{B'}{B} + \left(\frac{A'}{A} - \frac{B'}{B}\right)\frac{R'}{R}\right], \label{14c}\\
q &=& \frac{2}{AB}\left(\frac{\dot{R'}}{R} - \frac{\dot B}{B}\frac{R'}{R} - \frac{\dot R}{R}\frac{A'}{A}\right)
. \label{14d}
\end{eqnarray} This is an underdetermined system of four coupled partial differential
equations in seven unknowns viz., $ A, B, R, \mu, P_r, P_\perp $ and $q$.

\section{Exterior spacetime and junction conditions}

The exterior spacetime is taken to be Vaidya's outgoing solution,
given by \cite{vaidya}
\begin{equation} \label{v1}
ds^2 = - \left(1 - \frac{2m(v)}{R}\right) dv^2 - 2dvdR + R^2
\left(d\theta^2 + \sin^2\theta d\phi^2 \right)\,,
\end{equation}  where $m$($v$) represents the Newtonian
mass of the gravitating body as measured by an observer at
infinity. The necessary conditions for the smooth matching of the interior
spacetime (\ref{metric}) to the exterior spacetime (\ref{v1}) have been extensively investigated. We present the main results that are necessary for modeling a radiating star. The continuity of the intrinsic and extrinsic curvature components of the interior and exterior spacetimes across a time-like boundary are
\begin{eqnarray} \label{j}
m(v)_{\Sigma} &=& \left\{\frac{R}{2}\left[\left(\frac{\dot R}{A}\right)^2 - \left(\frac{R'}{B}\right)^2 + 1\right]\right\}_\Sigma \label{j1}\\
(P_r)_\Sigma &=& q_{\Sigma}\label{j2}.\end{eqnarray}
Relation (\ref{j2}) determines the temporal evolution of the collapsing star.

\section{Radiating Euclidean stars}

Following Herrera {\em et al}\cite{her1} we impose the condition that the areal radius of any spherical surface contained within $\Sigma$, with centre placed at the origin,  is equal to the proper radius from the center through to $r = b$, the boundary of the star.
This implies that
\begin{equation}
B = R' \label{euclid}
\end{equation}
The Einstein field equations (\ref{14a})--(\ref{14d}) reduce to
\begin{eqnarray} \label{eu}
\mu &=& \frac{1}{A^2}\left(\frac{\dot R}{R} + 2\frac{\dot R'}{R^{\prime}}\right)\frac{\dot R}{R}, \label{eu1}\\
P_r &=& -\frac{1}{A^2}\left[2\frac{\ddot R}{R} - \left(2\frac{\dot
A}{A} - \frac{\dot R}{R}\right)\frac{\dot{R}}{R}\right] +
2\frac{A'}{A}\frac{1}{RR'},\label{eu2}\\
P_{\perp} &=& -\frac{1}{A^2}\left[\frac{{\ddot R}}{R} + \frac{\ddot
R'}{R^{\prime}} - \frac{\dot A}{A}\frac{\dot R}{R} -
\left(\frac{\dot A}{A} - \frac{\dot R}{R}\right)\frac{\dot
R'}{R^{\prime}}\right]   \nonumber
\\&&+
\frac{1}{R'^2}\left[\frac{A''}{A} - \left(\frac{R''}{R^{\prime}} - \frac{R'}{R}\right)\frac{A'}{A}\right], \label{eu3}\\
q &=& -\frac{2}{AR'}\left(\frac{\dot{R}}{R}\frac{A'}{A}\right). \label{eu4}
\end{eqnarray}
and the mass function becomes
\begin{equation} \label{mass}
m = \frac{R}{2}\left(\frac{\dot R}{A}\right)^2\end{equation}
The boundary condition (\ref{j2}) yields
\begin{equation} \label{master}
\frac{\ddot R}{R} + \frac{1}{2}\left(\frac{\dot R}{R}\right)^2 - \frac{\dot A}{A}\frac{\dot R}{R} - \frac{(A + \dot{R})A'}{RR'} = 0
\end{equation}
valid on $r=b$.  We now have a system of six coupled partial differential equations, viz. (\ref{euclid})--(\ref{eu4}) and (\ref{master}), in seven unknowns.

We focus on (\ref{master}) as a solution of this equation will yield all the relevant kinematical and physical quantities. In doing so, we note that we are requiring (\ref{master}) to hold for all $r$ and not just on the boundary $r=b$. Equation (\ref{master}) is a nonlinear partial differential equation in $A(r,t)$ and $R(r,t)$.  We could analyse it as a quasi--linear partial differential equation in $A(r,t)$ only.  Then the general solution of (\ref{master}) will reduce to a general function of the solutions of two ordinary differential equations.  Unfortunately, these equations are still difficult to solve.  As a result, we provide a simple solution to (\ref{master}) by setting
\begin{equation}\label{m1}
R = \alpha A.\end{equation}
(Note that this closes our system of partial differential equations as we now have seven equations in seven unknown functions.)
This assumption leads to
\begin{equation}\label{soln1}
R(r,t) = \left[C_1(r)e^{\lambda_1 t} + C_2(r)e^{\lambda_2t}\right]^2,
\end{equation} where
\begin{equation} \label{sq}
\lambda_1 = \frac{1+\sqrt{3}}{2\alpha} \qquad \lambda_2 = \frac{1-\sqrt{3}}{2\alpha},
\end{equation}
which is the general solution to the resulting form of (\ref{master}) with the assumption (\ref{m1}) valid for all $r$.
Utilising solution (\ref{soln1}), the Einstein field equations (\ref{14a})-(\ref{14d}) yield
\begin{eqnarray}
\mu &=& 4\alpha^2{e^{-4\lambda_2t}\left[e^{\frac{\sqrt{3}t}{\alpha}}\lambda_1C_1 + \lambda_2C_2\right]}\times \nonumber\\&&\times
\frac{3e^{\frac{2\sqrt{3}t}{\alpha}}\lambda_1C_1C_1' + 3\lambda_2C_2C_2' + \sqrt{3}e^{\frac{\sqrt{3}t}{\alpha}}\left[\lambda_1C_1C_2' + \lambda_2C_2C_1'\right]}{(e^{\frac{\sqrt{3}t}{\alpha}}C_1 + C_2)^6(e^{\frac{\sqrt{3}t}{\alpha}}C_1' + C_2')} \\
P_r &=& \frac{-4\alpha e^{-4\lambda_2t}\left(e^{\frac{\sqrt{3}t}{\alpha}}\lambda_1C_1 + \lambda_2 C_2\right)}{(e^{\frac{\sqrt{3}t}{\alpha}}C_1 + C_2)^5}\\
P_{\perp} &=& - \frac{\left[(3 + 2\sqrt{3})e^{\frac{\sqrt{3}t}{\alpha}}C_1^{2}C_1' + (3 - 2\sqrt{3})C_2^2C_2'\right]e^{-4\lambda_2t}}{(e^{\frac{\sqrt{3}t}{\alpha}}C_1 + C_2)^6(e^{\frac{\sqrt{3}t}{\alpha}}C_1' + C_2')}  \nonumber\\ &&-
\frac{e^{\frac{\sqrt{3}t}{\alpha}}C_1C_2\left[(9+2\sqrt{3})e^{\frac{\sqrt{3}t}{\alpha}}C_1' + (9 - 2\sqrt{3})C_2'\right]e^{-4\lambda_2t}}{(e^{\frac{\sqrt{3}t}{\alpha}}C_1 + C_2)^6(e^{\frac{\sqrt{3}t}{\alpha}}C_1' + C_2')}\\
q &=& -\frac{4\alpha e^{-4\lambda_2t}\left(\lambda_1e^{\frac{\sqrt{3}t}{\alpha}} C_1 + \lambda_2 C_2\right)}{(e^{\frac{\sqrt{3}t}{\alpha}}C_1 + C_2)^5}
\end{eqnarray}
where $\lambda_1$ and $\lambda_2$ are defined in (\ref{sq}).
The magnitude of the shear tensor is given by
\begin{equation} \label{shear}
\sigma =
\frac{\sqrt{3}e^{(-1+2\sqrt{3})t/\alpha}\left(C_2C_1^{\prime} -
C_1C_2^{\prime}\right)}{(e^{\frac{\sqrt{3}t}{\alpha}}C_1 + C_2)^3(e^{\frac{\sqrt{3}t}{\alpha}}C_1' + C_2')}.
\end{equation}
Relation (\ref{shear}) indicates that the shear vanishes when $C_1(r) \propto C_2(r)$. In the next section we study the thermodynamical properties of our model. In order to ensure that the shear remains finite and nonzero for all time we make the following choice for our metric function
\begin{equation}
R(r,t) = \left[(a^2 + r^2)e^{\lambda_1t}+ (c^2 + r^2)e^{\lambda_2t}\right]^2,
\end{equation}
where $a$ and $c$ are constants and the $\lambda_i$ were defined earlier.

\section{Thermodynamics}

In this section we investigate the evolution of the temperature
profile of our model within the context of extended irreversible
thermodynamics. The causal transport equation in the absence of
rotation and viscous stress is \begin{equation} \label{causalgen}
\tau h_a{}^b \dot{q}_b+q_a = -\kappa \left( h_a{}^b\nabla_b
T+T\dot{u}_a\right)
\end{equation} where $h_{ab}=g_{ab}+u_a
u_b$ projects into the comoving rest space, $T$ is the local
equilibrium temperature, $\kappa$ ($\geq0$) is the thermal
conductivity, and $\tau$ ($\geq 0$) is the relaxational time-scale
which gives rise to the causal and stable behaviour of the theory.
To obtain the noncausal Fourier heat transport equation we set
$\tau = 0$ in (\ref{causalgen}). For the metric (\ref{metric}),
equation (\ref{causalgen}) becomes \begin{equation} \label{causal}
\tau{(qB)}\!\raisebox{2mm}{$\cdot$}  + AqB = -\frac{\kappa
(AT)'}{B}\,.
\end{equation}
In order to obtain a physically reasonable stellar model we will
adopt the thermodynamic coefficients for radiative transfer. Hence
we are considering the situation where energy is transported away
from the stellar interior by massless particles, moving with long
mean free path through matter that is effectively in hydrodynamic
equilibrium, and that is dynamically dominant. Govender {\em et
al}\cite{{gov1},{gov2}} have shown that the choice
\begin{equation} \kappa =\gamma T^3{\tau}_{\rm c}, \hspace{2cm}
 \tau_{\rm c} =\left({\alpha\over\gamma}\right)
T^{-\sigma}, \hspace{2cm}\tau =\left({\beta\gamma \over
\alpha}\right) \tau_{\rm c} \,, \label{5c}\end{equation} is a
physically reasonable choice for the thermal conductivity
$\kappa$, the mean collision time between massive and massless
particles $\tau_c$ and the relaxation time $\tau$. The quantities
$\alpha \geq 0$, $\beta \geq 0$ and $\sigma \geq 0$ are constants.
Note that the mean collision time decreases with growing
temperature as expected except for the special case $\sigma = 0$,
when it is constant.   With these assumptions the causal heat
transport equation (\ref{causal}) becomes
\begin{equation} \label{mm}
\beta(qB)\!\raisebox{2mm}{$\cdot$} T^{-\sigma} + A(qB) =
-\alpha\frac{T^{3 -\sigma}(AT)'}{B}\,.\end{equation} This equation was comprehensively studied in the noncausal  $(\beta = 0)$ case by Govinder and Govender\cite{gov3} as well as in specific causal cases.
In the noncausal case, (\ref{mm}) can be solved to yield
\begin{eqnarray}
(A{\tilde T})^{4 - \sigma} &=& \frac{\sigma - 4}{\alpha}\int{A^{4
- \sigma}qB^2dr} + F(t), \hspace{2cm} \sigma \neq 0 \\ \nonumber \\
\ln{(A{\tilde T})} &=& -\frac{1}{\alpha}\int{qB^2 dr} + F(t),
\hspace{2cm} \sigma = 4\,. \end{eqnarray} where $F(t)$ is a
function of integration which is fixed by the surface temperature
of the star. Note that $\tilde {T}$ corresponds to the noncausal
temperature when $\beta = 0$. For a constant mean collision time
$(\sigma = 0)$, (\ref{mm}) can be integrated to give the causal
temperature, ie.,
\begin{equation}
(AT)^4 =
-\frac{4}{\alpha}\left[\beta\int{A^3B(qB)\!\raisebox{2mm}{$\cdot$}
dr} + \int{A^4qB^2dr}\right] + F(t)\,.\end{equation} In
(\ref{5c}) we can think of $\beta$ as the `causality' index,
measuring the strength of relaxational effects, with $\beta=0$
giving the noncausal case.

The effective surface temperature of a star is given by
\begin{equation}
({\bar T}^4)_{\Sigma} =
\left(\frac{1}{r^2B^2}\right)\left(\frac{L}{4\pi\delta}\right)\,,\end{equation}
where $L$ is the luminosity at infinity and $\delta (>0)$ is a
constant. The luminosity at infinity can be calculated from
\begin{equation}
L_{\infty} = -\frac{dm}{dv}\,,\end{equation} where
$m(v)$ is given in (\ref{mass}).
We are in a position to analyse the evolution of the temperature in both the causal
and noncausal theories. Figure 1 represents the causal temperature (dashed line) and noncausal temperature (solid line) as function of the radial coordinate. It is clear that the temperature in both the causal and noncausal theories is a maximum at the center of the star and drops off smoothly as the radial coordinate increases towards the boundary. This trend also indicates that the surface layers of the star are much cooler than the interior portions. As in the acceleration-free case studied by Naidu {\em et al}\cite{nol} and Rajah and Maharaj\cite{rajah}, the causal temperature is everywhere higher than its noncausal counterpart at each interior point of the star. The causal and noncausal temperatures are equal at the boundary of the star. Figure 1 also reveals that relaxational effects account for a larger temperature gradient within the stellar core. This is expected at late times during the collapse as the stellar fluid is far from hydrostatic equilibrium. Figure 2 illustrates the trend in the relaxation times for the shear stresses. Following Naidu {\em et al}\cite{nol}, the shear transport equation yields
\begin{equation} \label{etaa}
{\tau}_1 = \frac{-P}{\dot{P} + \frac{8}{15}r_0\sigma
T^4},\end{equation} where coefficient of shear
viscosity for a radiative fluid
\begin{equation}\label{eta}
\eta = \frac{4}{15}r_0T^4\tau_1,\end{equation} was utilised. In (\ref{etaa}) we have used $P=
\frac{1}{3}\left(P_\perp - P_r\right)$ and $r_0$ is the
radiation constant for photons. We have further assumed that
$\tau_1 = \beta_1 \tau_c$. Figure 2 clearly shows that the relaxation time for the shear stresses can vary as much as a factor of $10^2$ during the evolution of the collapsing fluid.  A similar result was found for the acceleration-free model investigated by Naidu {\em et al}\cite{nol}. Figure 3 shows the proper radius as a function of time. It is a monotonically decreasing function as expected since the star is losing mass in the form of a radial heat flux. It is interesting to note that the formation of the horizon can be avoided in our model even in the presence of shear, by carefully choosing the arbitrary functions $C_1(r)$ and $C_2(r)$. Such a choice would ensure that the mass-to-radius ratio, $2 m_{\Sigma}/{\bar{r}_{\Sigma}} <
1$  which avoids the appearance of the horizon for all time. The horizon-free model of a radiating, shear-free star undergoing collapse was first studied by Banerjee {\em et al}\cite{hori}. The physical viability of this model was studied by Naidu and Govender\cite{nol1} where it was shown that the temperature and luminosity profiles were well behaved throughout the stellar interior.

In conclusion, we have presented an exact solution that completely describes the temporal and radial behaviour of a particular class of radiating stars, the so-called Euclidean stars. We have shown that the model is reasonably well-behaved throughout the collapse process, with the physical and thermodynamical variables remaining physically viable. Our model of a radiating star with nonvanishing shear adds to the very limited class of such solutions that are currently available in the literature.

\newpage
\begin{figure}[t]
\centerline{ \psfig{figure=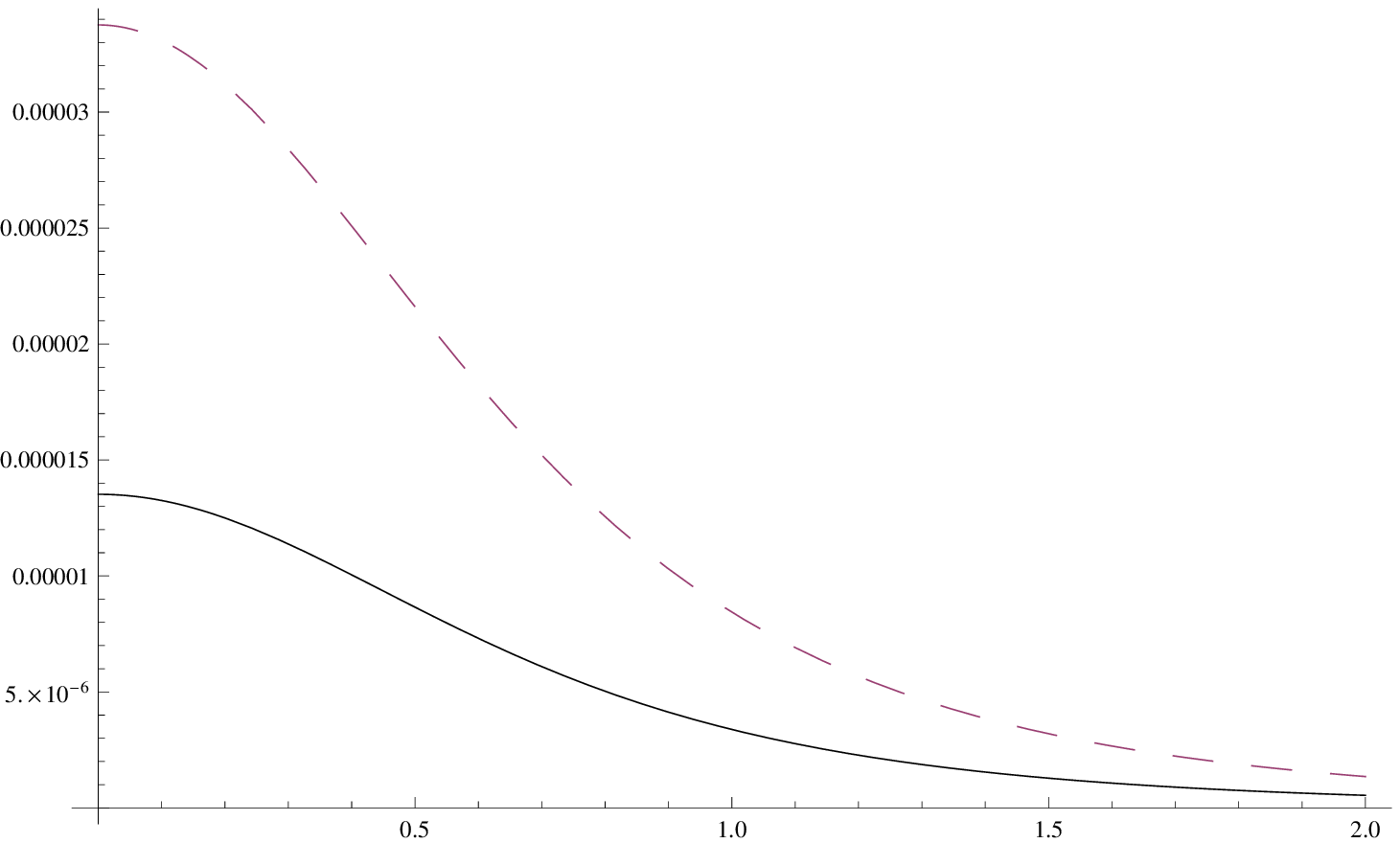}}
\caption{Causal (dashed line)and noncausal (solid line) temperature profiles versus $r$.}
\label{fig1}
\end{figure}
\begin{figure}[t]
\centerline{ \psfig{figure=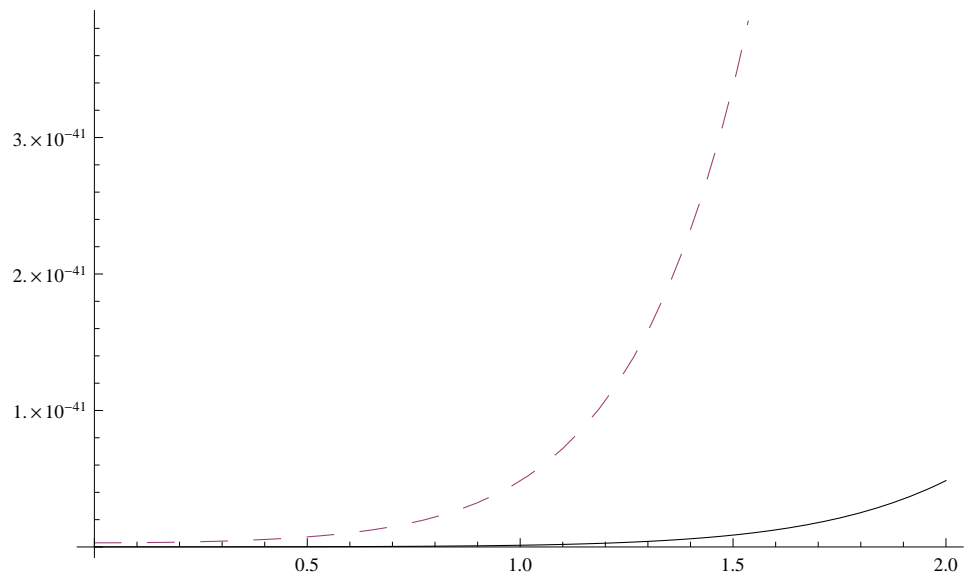}}
\caption{Relaxation time for the shear stress (close to equilibrium
- dashed line), (far from equilibrium - solid line) versus $r$.}
\label{fig2}
\end{figure}
\begin{figure}[t]
\centerline{ \psfig{figure=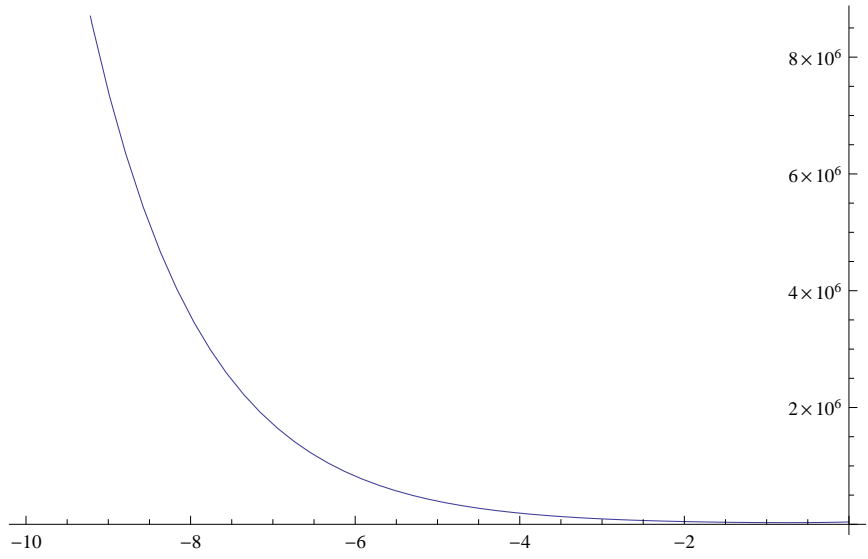}}
\caption{Proper radius versus time.}
\label{fig3}
\end{figure}

\end{document}